\documentclass[letterpaper]{article}

\usepackage[T1]{fontenc}

\usepackage{geometry}
\usepackage{setspace}

\usepackage{achemso}
\setkeys{acs}{articletitle = true}

\usepackage{graphicx}
\usepackage{float}
\newfloat{scheme}{htbp}{los}
\floatname{scheme}{Scheme}
\floatname{chart}{Chart}
\newfloat{graph}{htbp}{loh}

\usepackage{chemformula} 
\usepackage[version = 4]{mhchem} 

\usepackage{booktabs}
\usepackage{bm}
\usepackage{url}
\usepackage{array} 
\usepackage{longtable} 
\usepackage{hyphenat}
\usepackage{placeins}
\usepackage{amsmath}
\usepackage{amssymb}
\usepackage{bm}
\usepackage{comment}
\usepackage{wrapfig}

\newcommand{\tn}[1]{\textnormal{#1}}  

\renewcommand{\epsilon}{\varepsilon}
\graphicspath{}

\usepackage{authblk}
\author[1]{Vid Ravnik}
\author[2]{Baptiste Chabaud}
\author[1,3,4]{Urban Bren}
\author[2]{Galina V. Dubacheva*}
\author[5,6,7]{Tine Curk*}
\affil[1]{Faculty of Chemistry and Chemical Engineering, University of Maribor, Smetanova ulica 17, 2000 Maribor, Slovenia}
\affil[2]{Département de Chimie Moléculaire, Université Grenoble Alpes, CNRS UMR 5250, 570 rue de la chimie, CS 40700, 38000 Grenoble, France}
\affil[3]{The Faculty of Mathematics, Natural Sciences and Information Technologies, University of Primorska, Glagoljaška ulica 8, 6000 Koper, Slovenia}
\affil[4]{Institute for Environmental Protection and Sensors, Beloruska ulica 7, 2000 Maribor, Slovenia}
\affil[5]{Department of Materials Science and Engineering, Johns Hopkins University, 3400 North Charles Street, Baltimore, Maryland 21218, United States}
\affil[6]{Department of Physics and Astronomy, Johns Hopkins University, 3400 North Charles Street, Baltimore, Maryland 21218, United States}
\affil[7]{Institute for NanoBioTechnology, Johns Hopkins University, 3400 North Charles Street, Baltimore, Maryland 21218, United States}

\title{Kinetic Superselectivity in Multivalent Binding}
\date{*Email: tcurk@jhu.edu, galina.dubacheva@univ-grenoble-alpes.fr}
\begin{document}


\maketitle

\begin{center}
\textbf{Abstract}
\end{center}

\small

\begin{wrapfigure}{r}{0.54\textwidth}
    \centering
    \vspace{-10pt}
    \includegraphics[width=0.52\textwidth]{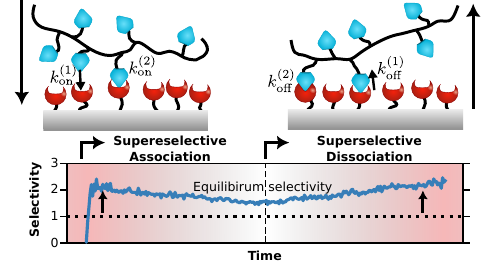}
    \vspace{-10pt}
\end{wrapfigure}

Multivalent binding employs multiple simultaneous supramolecular interactions, increasing avidity and selectivity compared with monovalent binding. While equilibrium aspects of multivalency are well characterized, non-equilibrium behavior remains poorly understood. By combining experiments on hyaluronic acid polymers with kinetic modeling based on stochastic chemical kinetics and molecular dynamics simulations, we systematically investigate the kinetics of multivalent binding. Notably, we find that both association and dissociation kinetics can be more selective than equilibrium binding. We explain this behavior using a two-step binding model featuring a combination of fast, weak and slow, strong interactions. These findings demonstrate a new approach: superselective targeting based on the association rate instead of the equilibrium state. The kinetic theory and experiments presented here provide a fundamental understanding of multivalent kinetics and establish design rules for superselective targeting in out-of-equilibrium systems.

\normalsize
\vspace*{1em}






%
\ Multivalency is characterized by the simultaneous interaction of multiple binding partners~\cite{mammen1998polyvalent}. 
While multivalent interactions result in increased binding strength (avidity) as compared to their monovalent counterparts, a more interesting consequence 
is superselectivity~\cite{martinez2011designing}.
Due to the combinatorial entropy of multiple bound states, multivalent binders can sharply discriminate between surfaces based on their receptor concentration or other control parameters, facilitating supramolecular recognition and selective targeting applications~\cite{Carlson2007,Wang2020,tian2020design,woythe2021quantitative,dubacheva2023determinants}.  

The thermodynamics of multivalent binding has been the subject of numerous theoretical and experimental studies~\cite{kitov2003nature,diestler2008statistical,martinez2011designing,dubacheva2023determinants,csizmar2018multivalent,li2025multivalent}, 
and the design principles for multivalent superselectivity in equilibrium are well understood. 
However, practical synthetic and biological systems often rely on transient encounters under transport and turnover constraints, 
so the relevant question is not only how a multivalent entity binds at equilibrium, but also how selectively it accumulates, persists, or is cleared. 
Kinetic selectivity may thus be critical for processes such as cell (pathogen) adhesion and communication~\cite{mulder2004multivalency,long2006kinetics,munoz2013real,an2022insights} and immune recognition~\cite{huang2010kinetics,liu2014accumulation,wu2019mechano}, where early binding events can trigger downstream responses. 
Understanding kinetic selectivity can also guide the design of biosensors or multivalent therapeutics that must achieve rapid and selective capture in complex environments, while avoiding off-target binding~\cite{stylianopoulos2015design,foster2017improved,shao2023review,mccann2025review}. 
%

\begin{figure}[htb!]
    \centering

    \includegraphics[width=3.3in]{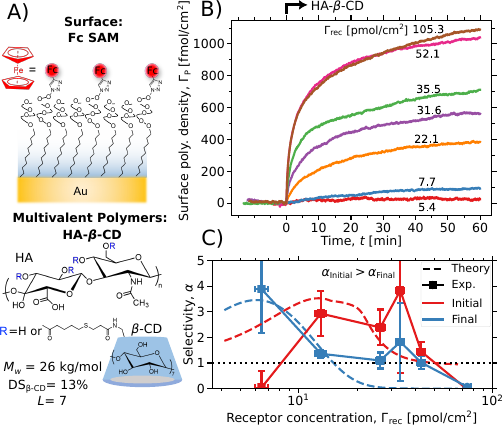}
    
    \caption{Experimental association kinetics of multivalent polymers.
    A) The experimental HA-$\beta$-CD--Fc SAM system. 
    B,C) Surface polymer density~$\Gamma_\mathrm{P}$ determined by SPR and the associated final (blue) and initial (red) selectivity determined by SPR at different Fc surface densities. 
    Electrochemical quantification of Fc density~$\Gamma_\mathrm{rec}$ and SPR data acquisition and treatment are described in detail in SI section~1.  
    Dashed curves in B,C represent a two-step theory fit of the experimental data, discussed later in main text. 
    }
    \label{fig:Exp_assoc}
\end{figure}

Experimental investigations of multivalent kinetics~\cite{lanfranco2019kinetics,schulte2022multivalent} have consistently demonstrated that increased avidity observed in multivalent binding 
is primarily due to decreased dissociation rates~\cite{hong2007binding,tassa2010binding,munoz2013real,choi2013dendrimer,scheepers2020multivalent,hoogerheide2026logarithmic}. 
However, existing kinetic models explore at most trivalent interactions~\cite{muller1998model,tassa2010binding,vauquelin2013exploring,errington2019mechanisms,bruncsics2022mvsim} and do not explore superselective binding.
Our goal is to understand the kinetic response of multivalent interactions and 
predict selectivity during the association and dissociation processes. 
To this end, we develop a theoretical model of multivalent binding kinetics 
and compare its predictions to a well-defined experimental platform based on host-guest interactions between hyaluronic acid (HA) substituted with $\beta$-cyclodextrin ligands~\cite{dubacheva2023determinants, chabaud2024influence} and self-assembled monolayers (SAMs) functionalized with ferrocene (Fc). 
We further validate the kinetic model with molecular dynamics (MD) simulations.


\ Using the experimental model system (Figure~\ref{fig:Exp_assoc}A), we investigate the kinetic response upon injection of multivalent polymers to surfaces with tunable receptor (Fc) densities~$\Gamma_\mathrm{rec}$. 
The surface chemistry and synthetic protocols for Fc SAM formation of and HA functionalization with $\beta$-CD are described in our previous studies~\cite{dubacheva2015designing,chabaud2024influence} and SI section~1. Specificity of host/guest interactions is confirmed by quartz crystal microbalance with dissipation monitoring (QCM-D) (SI Figure~S2).
Each sample is characterized by cyclic voltammetry to quantify Fc density (SI Figure S3), followed by in situ surface plasmon resonance (SPR) characterization to determine adsorbed HA-$\beta$-CD ($\beta$-CD - ligands) concentration versus time $\Gamma_\mathrm{P}(t)$ (Figure~S4). 
The obtained results are summarized in Figure~\ref{fig:Exp_assoc}B. 

To quantify the selectivity of binding under non-equilibrium conditions, we generalize the equilibrium~\cite{martinez2011designing,dubacheva2023determinants} definition of selectivity~$\alpha$ to depend on observation time $t$,
\begin{equation}    \label{eq:alpha}
    \alpha(t) = \frac{\mathrm{d}\ln\Gamma_\mathrm{P}(t)}{\mathrm{d}\ln\Gamma_\mathrm{rec}}  \;.
\end{equation}
Values of $\alpha$ above unity represent superselectivity, a superlinear dependence of binding on a control parameter, here $\Gamma_\mathrm{rec}$.
We calculate the time-dependent selectivity $\alpha(t)$ (SI Figure~S5) and define two limiting cases: the initial selectivity $\alpha_\mathrm{Initial}$ 
and final selectivity $\alpha_\mathrm{Final} = \alpha(t_\mathrm{Final})$ determined at the last data point, which is close to thermodynamic equilibrium and therefore analogous to the Hill coefficient~\cite{martinez2011designing,curk2018_ch3}. 
To reduce statistical noise, we define $\alpha_\mathrm{Initial}$ as the selectivity observed when the surface coverage reaches a predefined fraction ($1/4$) of the equilibrium (final) value (an alternate definition based on time yields very similar results, see SI Figure~S6). 

Surprisingly, the experimental system exhibits superselectivity in the initial association  ($\alpha_\mathrm{Initial}>1$), which significantly exceeds the final, equilibrium selectivity (Figure~\ref{fig:Exp_assoc}C) for a wide range of receptor densities.  The same trend is observed by reanalyzing published experimental data on a similar host/guest system with larger HA-$\beta$-CD multivalent probes~\cite{dubacheva2015designing} (SI Figure~S9).

\begin{figure}
    \centering
    \includegraphics[width=3.3in]{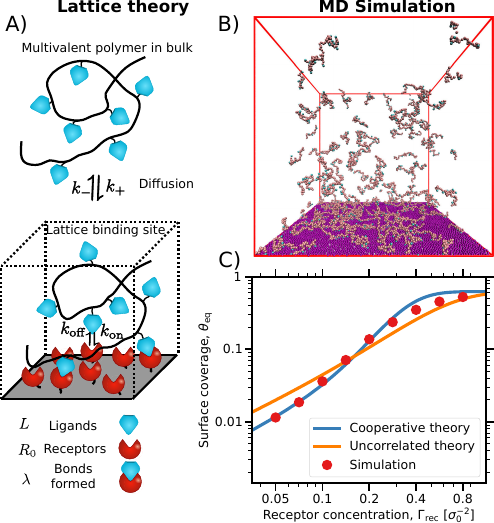}

    \caption{Kinetic theory and MD simulations. 
    A) Schematic of the developed theory, multivalent polymers adsorb from bulk solution into surface lattice sites with receptors.
    B) Example snapshot of MD simulations, multivalent polymers adsorb onto a receptor covered surface.
    C) Comparison of equilibrium (final) surface coverage, $\theta_\mathrm{eq}$ between MD simulations (red points) and a one parameter ($\epsilon(1)$) theory fit with uncorrelated binding ($\epsilon(\lambda) = \epsilon(1)$, orange line) and cooperative polymer binding (Eq.~\eqref{eq:epslam}, blue line).
    }
    \label{fig:theory_sim}
\end{figure}

To explain these observations, we propose a Langmuir-like model of multivalent kinetics based on previous thermodynamic theory~\cite{martinez2011designing,dubacheva2015designing}. We first investigate one-step binding, where ligand--receptor bond formation occurs in a single reaction step. 
Multivalent entities (e.g. polymers, nanoparticles)  with valency $L$ 
adsorb from bulk solution onto a receptor covered surface divided into lattice binding sites of volume $a^3$, for polymers $a^3 = 4\pi R_\mathrm{g}^3/3$, with $R_\mathrm{g}$ the polymer radius of gyration~\cite{dubacheva2014superselective}, 
and each site contains $R_0$ surface receptors (Figure~\ref{fig:theory_sim}A). 
We model multivalent kinetics with a system of ordinary differential equations (ODEs) based on stochastic chemical kinetics~\cite{mcquarrie1967stochastic,gillespie2007stochastic}. $\Gamma_\lambda$ denotes the concentration of surface sites that contain a multivalent entity with $\lambda$ formed bonds, and its evolution is described by
\begin{align}
    \begin{split}
    \frac{\mathrm{d}\Gamma_{\lambda}}{\mathrm{d}t} &= 
    k_\mathrm{off}(\lambda +1)\Gamma_{\lambda+1}
    - k_\mathrm{on,\lambda+1}(L-\lambda)(R_0-\lambda)\Gamma_\lambda  \\ 
    &- k_\mathrm{off}\lambda\Gamma_{\lambda} 
    + k_\mathrm{on,\lambda}(L-\lambda+1)(R_0-\lambda+1)\Gamma_{\lambda-1}  \;.
    \end{split}
    \label{eq:dGdt}
\end{align}
The ligand--receptor off rate~$k_\mathrm{off}$ is a constant, while the on rate~$k_\mathrm{on,\lambda}$ can depend on the number of formed bonds due to cooperative binding. 
For a multivalent entity with no bonds, $\lambda=0$, the last two terms in Eq.~\eqref{eq:dGdt} are replaced with terms for multivalent entity diffusion between surface and bulk solution, $-k_-\Gamma_0+k_+c_\mathrm{M}\Gamma_\mathrm{S}$, where $k_+ = Da$ and $k_- = D/a^2$ are the forward and reverse diffusive kinetic constants, with $D$ the diffusion constant of the multivalent entity, $\Gamma_\mathrm{S}$ is the concentration of empty surface binding sites and $c_\mathrm{M}$ the bulk multivalent entity concentration (Figure~\ref{fig:theory_sim}A).
We solve the system of ODEs to calculate the time evolution of the surface coverage $\theta(t) = \sum_{\lambda=1}^{\min(L,R_0)}\Gamma_\lambda(t)a^2$, and obtain results for a given receptor concentration~$\Gamma_\mathrm{rec}$, via a (Poisson) distribution of $R_0$. Surface coverage is proportional to the polymer molar density obtained from experiment, $\theta(t) = \Gamma_\mathrm{P}(t) N_\mathrm{A} a^2$, with $N_\mathrm{A}$ the Avogadro's constant.  Additional details can be found in SI section~2.

Previous thermodynamic models of multivalent binding have mostly considered the free energy of bond formation to be a constant. Multivalent nanoparticle binding can be described relatively well by this assumption~\cite{martinez2011designing,curk2017optimal}, however, multivalent polymers exhibit cooperative effects. 
Ligand binding brings the polymer backbone closer to the surface, which facilitates further ligand binding and increases selectivity~\cite{dubacheva2015designing,ravnik2024designing}. 
To calculate these effects we derive a scaling theory for cooperative polymer binding. The volume that unbound ligands can explore depends on the average loop size of the bound polymer backbone, and the on-rate is inversely proportional to this free ligand volume. The radius of gyration of the average loop is $R_\mathrm{g,loop}(\lambda)\sim R_\mathrm{g}/\lambda^\nu$ and the free volume that unbound ligands can explore is $\sim (R_\mathrm{g,loop})^3$. Therefore, we obtain the on rate for $\lambda$-th bond formation $k_\mathrm{on,\lambda} = k_\mathrm{on}\lambda^{3\nu}$, with $k_\mathrm{on}$ the on rate for first bond formation and $\nu$ the polymer scaling exponent. Here we use $\nu=0.5$ because the polymer is semi-flexible. The free energy of $\lambda$-th bond formation~$\epsilon(\lambda)$ is determined by the ratio of the rates: $\beta\epsilon(\lambda)=-\ln(k_\mathrm{on,\lambda}/k_\mathrm{off})$, with $\beta=1/k_\mathrm{B}T$, $k_\mathrm{B}$ the Boltzmann constant and $T$ the absolute temperature. The bond strength thus increases the more bonds are formed, 
\begin{equation}
\beta\epsilon(\lambda) = \beta\epsilon(1)-3 \nu\ln(\lambda)\;,
\label{eq:epslam}
\end{equation}
with $\epsilon(1)$ the free energy to form the first bond for an unbound polymer at the surface. 

%

The model is validated with MD simulations of bead-spring polymers adsorbing to a surface with the polymer bead diameter, $\sigma_0=1$~nm, and 48 beads per polymer to match the experimental contour length (Figure~\ref{fig:theory_sim}B, see methods, SI section~3 for details). 
The derived cooperative theory explains equilibrium MD results very well, significantly better than a description with uncorrelated binding ($\epsilon(\lambda) = \epsilon(1)$), which ignores polymer cooperative effects (Figure~\ref{fig:theory_sim}C).

\begin{figure*}[htb]
    \centering

    \includegraphics[width=\textwidth]{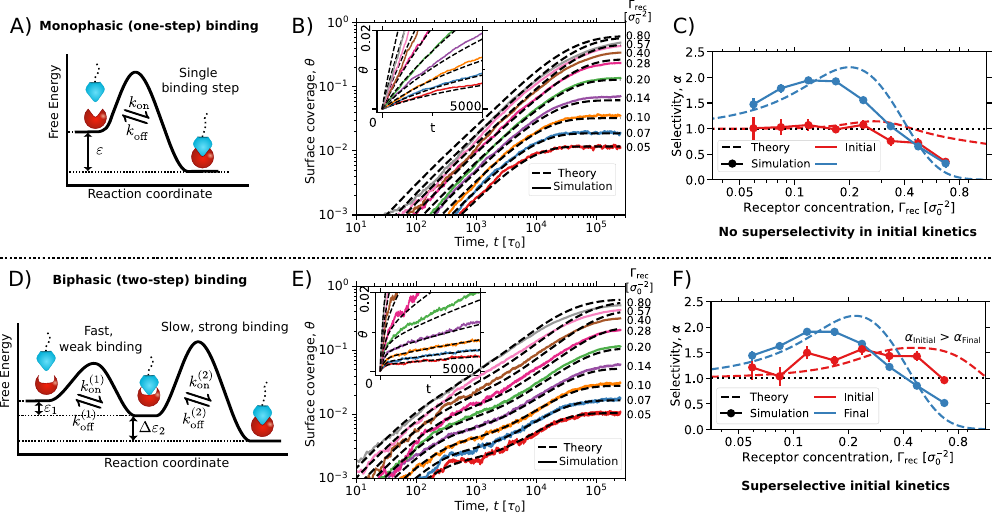}
    \caption{Modeling multivalent kinetics.
    A) Reaction barrier schematic for one-step (monophasic) binding. 
    B,C) Surface coverage and associated selectivity obtained from one-step MD simulations (full lines with markers)  and theoretical model (dashed lines). 
    D) Two-step (biphasic) binding features formation of an intermediate complex followed by a slower reaction to a more stable complex. 
    E,F)  Surface coverage and the associated selectivity for two-step MD simulations. 
    Insets in B,E show a zoom in at small surface coverages on a linear plot. 
    Dotted black lines in C,F represent selectivity $\alpha=1$, above which lies the superselective regime. 
   Theory fitted to one-step MD data ($\theta(t)$) with two fitting parameters ($\epsilon(1)$, $k_\mathrm{on}$), and to two-step MD data with four parameters, ($\epsilon(1)$, $k^{(1)}_\mathrm{on}$, $\Delta\epsilon_2$, $k^{(2)}_\mathrm{on}$).  
    }
    \label{fig:MD_assoc}
\end{figure*}

Cooperative theory and MD simulations also show very good agreement in adsorption kinetics (Figure~\ref{fig:MD_assoc}B,C), with deviations occurring only at very high receptor density ($\Gamma_\mathrm{rec} \gtrsim \sigma_0^{-2}$) where receptor crowding reduces the association rate (Figure~\ref{fig:MD_assoc}B).  
Additionally, Eq.~\eqref{eq:epslam} slightly overestimates polymer cooperativity (See SI, section 4.1), leading to overestimation at strong binding. 
While this system exhibits superselectivity in equilibrium, the initial, kinetic selectivity in association dynamics is not superselective, $\alpha_\mathrm{Initial} \lesssim 1$ (Figure~\ref{fig:MD_assoc}C). This result can be intuitively understood because the association rate is determined by the formation of the first bond between the multivalent entity and surface receptors, the rate of which is proportional to the receptor concentration, so we expect $\alpha_\mathrm{Initial} \approx 1$ for a reaction-limited case (and $\alpha_\mathrm{Initial} \approx 0$ for the diffusion-limited case where association is independent of surface properties). 
This is supported by the fact that multivalent avidity effects arise primarily due to decreased unbinding rates~\cite{tassa2010binding,munoz2013real,choi2013dendrimer},  while the total reaction forward rate increases only linearly with the number of ligands~\cite{hong2007binding} (see SI section~4).
However, $\alpha_\mathrm{Initial} \approx 1$ is inconsistent with our experimental data (Figure~\ref{fig:Exp_assoc}C) which clearly show a superselective kinetic response.

As one-step multivalency fails to exhibit superselectivity in association kinetics, 
we investigate biphasic, two-step binding, 
commonly observed in various kinetics studies~\cite{tiwari2015analyzing}. The main feature of two-step binding is an initial, rapid formation of an intermediate complex, followed by slower formation of a more stable complex. Such binding is found in many supramolecular and biological systems, where the formation of the stable complex is usually accompanied by a slow conformational change in the receptor (induced-fit model)~\cite{koshland1995key}.
We extend the theoretical model to include two-step ligand--receptor binding (cf. Figures~\ref{fig:MD_assoc}A,D).
The microscopic on and off reaction rates between unbound ligands and receptors and the intermediate bound state are $k^{(1)}_\mathrm{on,\lambda}$, $k^{(1)}_\mathrm{off}$, while reaction rates between the intermediate and the stable state are denoted by $k^{(2)}_\mathrm{on}$, $k^{(2)}_\mathrm{off}$.  
The free energy of intermediate bond formation is $\epsilon_1(\lambda)$,  while the free energy change for forming the stable bond from the intermediate state is $\Delta\epsilon_2$, where $k^{(1)}_\mathrm{on,\lambda}/k^{(1)}_\mathrm{off} = \exp{(-\beta\epsilon_1(\lambda))}$ and $k^{(2)}_\mathrm{on}/k^{(2)}_\mathrm{off} = \exp{(-\beta\Delta\epsilon_2)}$. Cooperative effects depend on the total number of formed bonds, $\lambda=i+j$, analogously to the one-step model, see methods Eq.~\eqref{eq:dGdt_two_step}. 

We validate the two-step model via MD simulations, performing an identical set of MD simulations as before, changing only the ligand--receptor interaction potential to feature an additional higher barrier within the potential.  
Theory and MD show excellent agreement,  see Figure~\ref{fig:MD_assoc}E,F. 
Comparing the two-step and one-step cases, the equilibrium results are nearly identical, as thermodynamically, the two-step model is exactly equal to the monophasic model with a rescaled free energy of bond formation, $\beta\epsilon(\lambda) = -\ln\left[\exp({-\beta\epsilon_1(\lambda)})+\exp({-\beta(\epsilon_1(\lambda)+\Delta\epsilon_2)})\right]$. 
However, the kinetic response of two-step binding is superselective ($\alpha_\mathrm{Initial}>1$), and we observe a range where the initial selectivity is significantly higher than the equilibrium selectivity (Figure~\ref{fig:MD_assoc}F). This kinetic superselectivity is achieved on a range of timescales $t$ that fall between the formation of the weak intermediate bond and equilibration,
$ 1/(k_\mathrm{on}^{(1)}LR_0 c_\tn{M}a^3) \lesssim t \lesssim 1/k_\mathrm{on}^{(2)}$.

The kinetic superselectivity arises because the 
system initially reaches a multivalent pseudo-equilibrium dictated solely by the intermediate, fast interactions, before the binding slowly proceeds to the true equilibrium dictated by the slower, stable interactions. This pseudo-equilibrium features a selectivity maximum at a higher receptor concentration compared to the full equilibrium (Figures~\ref{fig:Exp_assoc}C,~\ref{fig:MD_assoc}F). 
As a result, there exists a range of receptor concentrations 
where the pseudo-equilibrium selectivity is larger than the equilibrium selectivity. 
%
%
Viewing the binding process through the rate-determining step (RDS), if the RDS occurs in the initial stages of the binding (diffusion, first bond formation) we cannot expect kinetic superselectivity, contrary to if the RDS occurs when binding is already multivalent.
This phenomenon is not limited to a two-step binding model, and we expect a qualitatively similar effect for any combination of multivalent weak, fast and slow, strong interactions (see SI Figure~S15). 
Intriguingly, even if the equilibrium binding is very strong (e.g. covalent), the system would still exhibit kinetic superselectivity. Thereby, superselective binding should not be limited to weak bonds as previously thought~\cite{martinez2011designing,dubacheva2023determinants,scheepers2020multivalent}. 
%

\begin{figure} 
    \centering
    \includegraphics[width=3.3in]{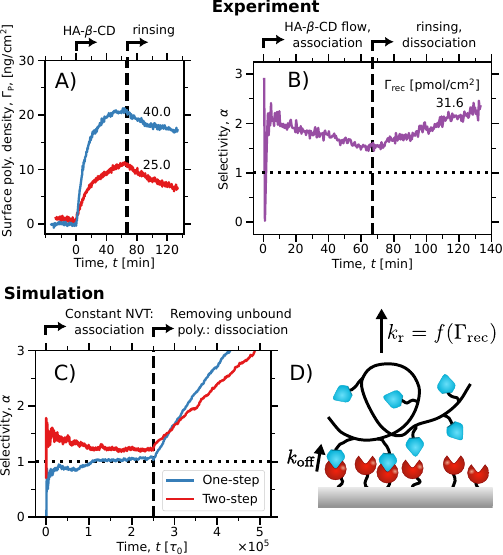}
    
    \caption{Selectivity in association and dissociation. 
    A,B) Surface density and the associated selectivity time dependence measured by QCM-D--SE. At $t=67$~min the polymer solution is replaced by its aqueous medium and dissociation begins. 
    C)  MD simulations selectivity time dependence for one-step (blue) and two-step (red) ligand--receptor binding. 
    D) The total dissociation rate constant, $k_\mathrm{r}$, decreases with receptor concentration, resulting in increasing dissociation selectivity.
    }
    \label{fig:dissoc_exp_MD}
\end{figure}

While the two-step binding model of fast and slow interactions can explain the observed kinetic superselectivity (Figure~\ref{fig:Exp_assoc}C), the exact chemical mechanism responsible for these observations remains an open question.  
We speculate the two-step binding behavior originates from a weak, hydrophobic interaction between the modified polymer and Fc, while $\beta$-CD--Fc host--guest binding represents the strong binding step. 
The weak, nonspecific interactions may be mediated by the primarily hydrophobic linker between $\beta$-CD and HA (see Figure~\ref{fig:Exp_assoc}A). This is supported by observation of weak adsorption of HA-$\beta$-CD to a hydrophobic surface composed of purely alkyl SAM (SI Figure~S3C,D), suggesting similar interaction is present on hydrophobic Fc SAM. 
%

\ In addition to SPR experiments, which are limited by the maximum injection volume, we performed a set of QCM-D in situ coupled with spectroscopic ellipsometry (SE) experiments (Figures S7,S8) employing the same Fc SAM, HA-$\beta$-CD system to study longer dissociation kinetics (Figure~\ref{fig:dissoc_exp_MD}A). 
Interestingly, the dissociation process is also more selective than the equilibrium interaction in both experiments and MD simulations (Figures~\ref{fig:dissoc_exp_MD}B,C). 
We explain this phenomenon by approximating the dissociation binding curve as a first order exponential function $\theta(t) = \theta_\mathrm{eq}\exp(-k_\mathrm{r}t)$, where $\theta_\mathrm{eq}$ is the surface density at the beginning of the dissociation and $k_\mathrm{r}$ the effective unbinding rate constant for the multivalent entities. Because the binding is multivalent, $k_\mathrm{r}$ decreases with receptor density (Figure~\ref{fig:dissoc_exp_MD}D). This results in a linear increase in dissociation selectivity with time, $\alpha(t) = \alpha_\mathrm{eq}
-\frac{\mathrm{d} k_\mathrm{r}}{\mathrm{d}\ln\Gamma_\mathrm{rec}}t$, where $\frac{\mathrm{d} k_\mathrm{r}}{\mathrm{d}\ln\Gamma_\mathrm{rec}} <0\;$, in agreement with both experiment and MD simulation data.
The selectivity will thus be maximized when the surface is completely empty, $\theta\to0$, which is likely not very useful. However, if we ask what is the maximal selectivity at a given value of~$\theta$, it turns out the largest selectivity is achieved in a dissociation process. 
In the strong binding limit the surface is initially fully covered, $\theta_\mathrm{eq} \to 1$, and we can derive $k_\mathrm{r}\propto\Gamma_\mathrm{rec}^{-L}$ (SI section~2.5). 
Thereby, the maximal dissociation selectivity at a given $\theta$ is
\begin{equation} \label{eq:alpha_theta_c_dissoc_limit}
\alpha^\mathrm{max}(\theta) = -L\ln[\theta(t)] \;.
\end{equation}
Whereas the equilibrium and association selectivity are limited by valency~\cite{martinez2011designing,curk2018_ch3}, $\alpha_\mathrm{eq} \leq L$, the selectivity in dissociation can surpass $L$ for $\theta(t) < 0.3$. 

The multivalent equilibrium selective response depends on the master, scaling variable~\cite{martinez2011designing,dubacheva2015designing}, and our kinetic model also predicts identical behavior for different systems with an identical value of a scaling variable, $\gamma=R_0 k_\mathrm{on}/k_\mathrm{off}$, provided that the off rate, $k_\mathrm{off}$, is kept constant (and assuming an abundance of receptors, $R_0 \gg L$, see SI section~2.3). Thus, while the experimental and simulation results demonstrate  kinetic selectivity to receptor concentration, we expect the results to be general across other control parameters such as bond strength, valency, temperature, or presence of co-factors or competitors~\cite{curk2022controlling,dubacheva2023determinants}.

\begin{figure}
    \centering
    \includegraphics[width=3.3in]{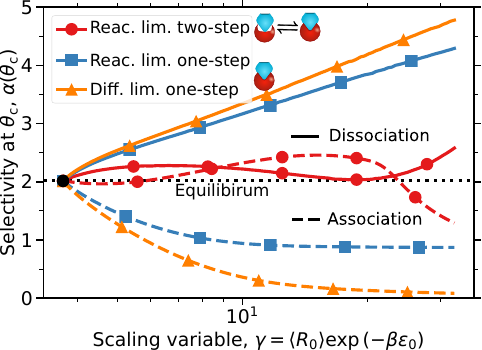}
   \caption{Selectivity at a given value of surface coverage, $\theta_\mathrm{c}$. 
   Black dotted line shows equilibrium selectivity, full, dashed lines show selectivity in dissociation, association kinetics respectively.     Blue squares represent reaction-limited ($D\to\infty$) monophasic binding, orange triangles show diffusion-limited monophasic binding ($D=10^3$~$\sigma_0^2\tau_0^{-1}$), and red circles shows reaction-limited two-step binding.  
   Parameters:   
   $\theta_\mathrm{c} = 0.25$,     
   $L = 5$, $R_\mathrm{g}=5~\sigma_0$, 
   $c_\mathrm{M}=10^{-6}~\sigma_0^{-3}$, uncorrelated binding, $\beta\epsilon(\lambda)=0$, $k_\mathrm{on,\lambda} = 1~\tau_0^{-1}$. 
   Two-step binding parameters:   
   $\beta\Delta\epsilon_2=-2$, $k^{(2)}_\mathrm{on}/k^{(1)}_\mathrm{on,1}=0.01$.    }
    \label{fig:alpha_at_theta}
\end{figure}

We demonstrate the behavior of selectivity at a given value of surface coverage, $\theta_\mathrm{c}$, by comparing $\alpha(\theta_\mathrm{c})$ as a function of the multivalent scaling variable $\gamma$ 
in association and dissociation kinetics, as well as when $\theta_\mathrm{c}$ is the equilibrium surface coverage (Figure~\ref{fig:alpha_at_theta}).  
As expected from the previous discussion, the highest selectivity at a given surface density $\theta_\mathrm{c}$ is obtained in dissociation kinetics. Association and dissociation show opposite behavior for both one-step, two-step and diffusion limited cases; when one is high, the other is low. We can understand the reversed effects by considering that the equilibrium value is obtained by the ratio of the association and dissociation rates, therefore, if selectivity in one increases it must decrease in the other, resulting in the mirroring between association/dissociation results around the equilibrium value.
Interestingly, the dissociation selectivity is higher when binding is diffusion-limited, compared to reaction-limited. Under diffusion limited conditions, polymers can rebind at the surface and the probability of rebinding depends on the scaling variable, thus increasing the dissociation selectivity (see SI section~2.5). 
Notably, in the two-step binding case, 
the association selectivity can exceed the equilibrium selectivity, however, this results in a corresponding decrease in dissociation selectivity. 
We expect similar behavior for other models exhibiting association kinetic supreselectivity, e.g. when ligands have two distinct binding modes. 



\ In conclusion, 
we identified pronounced superselectivity in association kinetics of multivalent polymers, which we explain using a two-step binding model that features a combination of fast, weak and slow, strong interactions. 
This finding introduces a new approach: superselective targeting based on the association rate rather than the equilibrium state, which should be very useful in systems where equilibrium is difficult to achieve. Moreover, this finding can be applied to introduce (kinetic) superselectivity to systems where equilibrium superselectivity is impossible, such as very strong (covalent) interactions.
We find that high kinetic selectivity in association is achieved on a timescale~$t$ that falls between the formation of the weak and strong bonds, for two-step binding  $1/(k_\mathrm{on}^{(1)}LR_0 c_\mathrm{M}a^3) \lesssim t \lesssim 1/k_\mathrm{on}^{(2)}$. 
Therefore, a large window of high kinetic selectivity is obtained when the rate of weak bond formation~$k_\mathrm{on}^{(1)}$ is much faster than the rate of strong binding~$k_\mathrm{on}^{(2)}$.


We demonstrated that selectivity in a dissociation process increases linearly with time, and surpasses the equilibrium selectivity. This leads to a surprising result that the maximal possible selectivity is achieved by fully saturating the surface and waiting for dissociation. 
In analogy with equilibrium predictions, we show that the kinetic selectivity is a general feature of multivalent interactions and extends to other control parameters beside receptor concentration. 

Practically, in the body, cells can endocytose macromolecular or nanoparticle based drugs on a timescale of seconds to minutes. If endocytosis time is comparable or faster than drug binding, then equilibrium is never achieved and therefore targeting selectivity is determined by the association rate selectivity, rather than equilibrium selectivity. On the other hand, many cellular processes require that a ligand is bound for a sufficient amount of time to trigger a response. In this case, targeting cells based on dissociation rate may yield optimal selectivity.  
More broadly, our results provide fundamental understanding of multivalent kinetics. The kinetic theory presented here should be useful to engineer selective multivalent interactions in biological or synthetic systems that operate out of equilibrium.

\FloatBarrier
\section*{Methods}

\ Self-assembled monolayer formation and functionalization with ferrocene through the azide-alkyne click reaction was accomplished using previously developed protocols~\cite{chabaud2024influence,3_dubacheva2010electrochemically}.  
HA-$\beta$-CD was synthesized by thiol–ene coupling between HA–pentenoate and $\beta$-CD–thiol~\cite{dubacheva2014superselective,chabaud2024influence}.  The HA-$\beta$-CD average molecular weight, the average distance between $\beta$-CDs along the contour of the polymer chain, and the average number of $\beta$-CDs per polymer chain were calculated to be $M_\mathrm{w}=26$~kg/mol, $d_\mathrm{\beta-CD}=7$~nm, and $L=7$ (degree of substitution, DS$_\mathrm{\beta-CD}=0.13$). 
Cyclic voltammetry was used to quantify Fc surface density in the formed SAM-Fc. Electrochemical measurements were carried out following previously established procedures using a three-electrode potentiostatic system~\cite{3_dubacheva2010electrochemically}.  
The areal density of Fc was modulated ($\Gamma_\mathrm{rec}$), and the binding response was studied via surface plasmon resonance (SPR) upon the injection of a fixed concentration $c_\mathrm{M}=50~\mu$g/mL of HA-$\beta$-CD.  
Over extended time intervals, quartz crystal microbalance with dissipation monitoring (QCM-D) enabled real-time monitoring of HA-$\beta$-CD binding and its subsequent desorption during rinsing. Spectroscopic ellipsometry was in situ coupled with QCM-D to monitor the adsorption of HA-$\beta$-CD and to establish a calibration curve correlating QCM-D frequency shifts with the corresponding adsorbed dry mass~\cite{4_kirichuk2023competitive}.  
Additional details are available in the Supporting Information section~1.


\ The developed theoretical model of multivalent binding kinetics is described in the main text Eq.~\eqref{eq:dGdt}, the derived expression for polymer cooperativity is given by Eq.~\eqref{eq:epslam}. 
The extended two-step model (Figure~\ref{fig:MD_assoc}D) tracks the concentration of states with $i$ intermediate bonds, and $j$ stable bonds, $\Gamma_{i,j}$,
\begin{align}
    \begin{split}
        \frac{\mathrm{d}\Gamma_{i,j}}{\mathrm{d}t} &=
    (i+1)k^{(1)}_\mathrm{off}\Gamma_{i+1,j}
    - k^{(1)}_\mathrm{on,\lambda+1}(L-\lambda)(R_0-\lambda)\Gamma_{i,j} \\
    &-i k^{(1)}_\mathrm{off}\Gamma_{i,j} 
    + k^{(1)}_\mathrm{on,\lambda}(L-\lambda+1)(R_0-\lambda+1)\Gamma_{i-1,j}\\
    &+ (j+1)k^{(2)}_\mathrm{off}\Gamma_{i-1,j+1} 
    - ik^{(2)}_\mathrm{on}\Gamma_{i,j} \\
    &- jk^{(2)}_\mathrm{off}\Gamma_{i,j}
    + (i+1)k^{(2)}_\mathrm{on}\Gamma_{i+1,j-1}     
    \;,
    \end{split}
    \label{eq:dGdt_two_step}
\end{align}
%
with $\lambda=i+j$. 
$\Gamma_{0,0}$ includes the diffusive contribution ($-k_-\Gamma_{0,0}+k_+c_\mathrm{M}\Gamma_\mathrm{S}$), while terms not originating from or leading towards valid states ($i+j \leq \min(L,R_0)$, $i,j\geq0$) are omitted.  
We include cooperative effects in the formation of the intermediate bonds, $k_\mathrm{on,\lambda}^{(1)} = k_\mathrm{on,1}^{(1)} \lambda^{3\nu}$, based on the total number of formed bonds, $\lambda=i+j$.
The Supporting information section~2 contains additional detail about the kinetic and associated thermodynamic equilibrium model, derivation of scaling variable, and additional comparison between model and experimental results.  
%
Theory fit parameters to experimental data in Figure~\ref{fig:Exp_assoc}C are: $\beta\epsilon(1) = 3.96$, $\beta\Delta\epsilon_2 = -0.56$, $k_\mathrm{on,1}^{(1)} =2.20\times10^{-3}$~s$^{-1}$, $k_\mathrm{on}^{(2)} = 2.72\times10^{-4}$~s$^{-1}$. In Figure~\ref{fig:MD_assoc}B,C  to one-step MD data the fit parameters are given by: 
 $\beta\epsilon(1) = 4.79$, $k_\mathrm{on,1} = 1.71\times10^{-6}~\tau_0^{-1}$,
 and in Figure~\ref{fig:MD_assoc}E,F the fit parameters to two-step MD data are: 
 $\beta\epsilon(1) = 4.88$, $\beta\Delta\epsilon_2 = -1.30$, $k_\mathrm{on,1}^{(1)} = 7.92\times10^{-6}~\tau_0^{-1}$, $k_\mathrm{on}^{(2)} = 2.72\times10^{-4}~\tau_0^{-1}$.

\ We employ LAMMPS~\cite{LAMMPS} to perform constant $NVT$ molecular dynamics (MD) simulations of a bead-spring multivalent polymer binding to a receptor-decorated surface. 
We use the polymer bead diameter, $\sigma_0$, ($\sigma_0=1$~nm) 
as the length unit scale of the simulations, while $\tau_0$  
is the time unit ($\tau_0=0.44$~ns). We simulate a cubic box $100~\sigma_0$ to a side with 100 flexible polymers consisting of 48 backbone beads and $L=8$ ligand beads randomly distributed along the backbone. The polymer concentration is $c_\mathrm{M}=10^{-4}$~$\sigma_0^{-3}$. We apply periodic boundary conditions in two spatial directions, while the third contains repulsive walls, one of which carries point like receptors which can interact with ligand beads on polymers. 
Ligand--receptor binding is described by a spherically symmetric potential which features an additional potential barrier in the case of two-step binding (cf. Figures~\ref{fig:MD_assoc}A,D, see SI Figure~S21). 
Receptors are stationary, not integrated during the production simulations, and are equilibrated via two-dimensional simulations with receptor size $\sigma_\mathrm{rec}=\sigma_0$ with WCA potential to prevent ligand binding to multiple receptors or vice versa.  
Before the association production runs, polymers were randomly distributed in the simulation box and equilibrated without the presence of receptors. 
Dissociation production runs began from the endpoints of association runs. We simulate a dissociative (zero bulk concentration) environment by periodically deleting any polymers above a height threshold 
from the simulation.
During production simulations, we track the number of formed ligand--receptor bonds, as well as the number of bound polymers. We average our results over a number of parallel simulations to reduce statistical noise.
Full details are provided in the Supporting Information section 3. To speed up MD simulations and enable exploration of relevant timescales, we use 100-fold larger polymer bulk concentration in the simulations compared to experiments. As a consequence, the selectivity in the experimental data is higher compared to the MD data (compare Figures~\ref{fig:Exp_assoc}C and~\ref{fig:MD_assoc}F), as higher concentration reduces selectivity~\cite{dubacheva2023determinants}.

\section*{Acknowledgements}
 V.R. and U.B. gratefully acknowledge financial support from the Slovenian Research and Innovation Agency (ARIS) through project and program grants P1-0403, P2-0438, J1-4398, and L7-60161. U.B. also acknowledges the Fulbright foundation for funding his research visit at the Johns Hopkins University. 
G.D. and B.C. acknowledge ANR JCJC ``SupraSwitch'' funding (ANR-18-CE09-0009) and ICMG FR2607 Chemistry Nanobio Platform (UGA, Grenoble). Angéline van Der Heyden (DCM, UGA, Grenoble) is acknowledged for her help with SPR experiments.
T.C. acknowledges support from the National Science Foundation, CBET Division, Biosensors Program, Award Number 2402407. 
Computational work was carried out at HPC Vega and the Advanced Research Computing at Hopkins (ARCH) core facility (rockfish.jhu.edu), which is supported by the National Science Foundation (NSF) grant number OAC 1920103. 

\section*{Supporting information}


The following files are available free of charge.
\begin{itemize}
  \item Additional details about the experiments (section~1),  the theoretical model of multivalent binding kinetics (section~2), molecular dynamics simulations (section~3), as well as further comparisons between simulation and theory results (section~4) (PDF). 
\end{itemize}

\bibliography{refs.bib}

\newpage


   

  
  

\end{document}